\documentclass[a4paper,11pt,openright,twoside]{article}
\usepackage[english,francais]{babel}
\usepackage[latin1]{inputenc}
\usepackage[T1]{fontenc}
\usepackage[dvips]{graphicx}

\usepackage[authoryear]{natbib}

\usepackage{url}
\def\auteurTitre{Bernard \textsc{Jacquemin}\up{1,2} et Sabine \textsc{Ploux}\up{1}}
\def\auteur{B. \textsc{Jacquemin} et S. \textsc{Ploux}}
 
\def\adresselabo{\up{1}ISC, Université Claude Bernard Lyon1, CNRS UMR5015\\
                 67, boulevard Pinel F-69\,675 Bron cedex\\
		 \up{2}ECA, Université de Haute Alsace Mulhouse, CNRS UMR7044\\
		 10, rue des Frères Lumière F-68\,093 Mulhouse cedex}
 
\def\courriel{\{Bernard.Jacquemin,sploux\}@isc.cnrs.fr}

\def\titre{Corpus spécialisé et ressource de spécialité: l'information forme le sens}

\def\titrecourt{Corpus spécialisé et ressource de spécialité}

\def\piedpage{Maniez, F., Dury, P., Arlin, N., Rougemont, C. (Dirs) (2008).\\\textit{Corpus et dictionnaires de langues de spécialité}, Grenoble, PUG, p.~197\textendash{}212.}

\usepackage{fancyhdr}
\title{\titre}
\author{\auteurTitre\\\adresselabo\\\courriel}
\date{}

\begin{document}

\maketitle

\begin{abstract}
\noindent Les Atlas sémantiques sont un modèle mathématique et statistique de représentation visuelle de la sémantique lexicale basé sur l'examen des relations entre les mots. Une application de ce modèle à des relations de proximité contextuelle dans un corpus a permis de montrer que le modèle était capable de dénoter le sens des unités lexicales tel qu'il est perçu par les rédacteurs du corpus. Nous nous appuyons sur ce constat pour proposer d'exploiter le modèle afin de construire automatiquement un dictionnaire spécialisé dans un domaine précis par l'analyse d'un corpus représentatif de ce domaine. Tout en conservant le modèle, nous modifions son application en faisant intervenir une analyse morphologique et syntaxique pour établir la réalité des unités lexicales ainsi que les liens entre elles, qui sont dès lors de nature syntaxique. Nous proposons également d'utiliser la ressource produite pour naviguer dans le corpus utilisé considéré comme une base d'information en suivant le sens plutôt que le mot. Enfin, nous proposons d'exploiter cette approche pour aider à la réalisation de dictionnaires plus classiques ou pour étudier la langue en diachronie. \\

%The "Atlas sémantiques" are a mathematical and statistical model that represents
\textbf{Mots-clefs:} Corpus spécialisé, dictionnaire spécialisé, gestion de l'information, clique, analyse factorielle des correspondances, sémantique lexicale, approche mixte, linguistique, statistique.
\end{abstract}

\section{Introduction}

Depuis quelques années, les disciplines liées à la gestion des connaissances (\textsl{knowledge management}) ont vu leur importance grandir à la fois dans les préoccupations scientifiques et dans les innovations technologiques. En effet, la maîtrise de l'information s'affirme comme un enjeu stratégique d'une importance majeure pour des domaines aussi divers que la politique, l'économie, la culture, l'innovation ou la recherche. Les thématiques du traitement de l'information textuelle sont bien entendu au c{\oe}ur de cette dynamique, traduite dans les faits par le succès de rencontres et compétitions telles que MUC (\textsl{Message Understanding Conference}) \citep{Chinchor92}, TREC (\textsl{Text Retrieval Conference}) \citep{Harman92,VoorheesBuckland05}, CLEF (\textsl{Cross-Language Evaluation Forum}) \citep{Peters02,PetersAl04} ou NTCIR (\textsl{NII Test Collection for IR Systems}) \citep{FukumotoKato01,KandoTakaku05}, ou par l'intérêt manifesté pour des approches de classifications sémantiques de pages Web, telles que les \textsl{Topic Maps} (\url{http://www.topicmaps.org/}),  et de contenus de pages Web, comme le \textit{Web Sémantique} (\url{http://www.w3.org/2001/sw/}). Toutefois, la difficulté qu'il y a à appréhender l'ensemble du monde, dans sa diversité, grâce à une base de connaissance unique, réaliste et cohérente, a favorisé l'éclosion d'approches en domaine restreint. On en vient donc à définir des sous-domaines de la langue, en limitant tant le lexique que les acceptions propres à ce lexique aux usages attestés dans la thématique considérée.

La problématique, qui consistait donc à établir un outil de représentation du monde et de navigation dans cette représentation, a donc évolué. Elle réside à présent dans la création de ressources partielles contenant les connaissances et les caractéristiques langagières propres au domaine considéré. Cependant, les ressources propres à un domaine sont rares, et elles sont d'autant plus incomplètes qu'un domaine est novateur ou dynamique. De ce fait, les domaines qui, de par leur activité, sont les plus demandeurs en système de gestion de l'information sont également ceux qui sont le moins susceptibles de disposer d'outils adaptés à leurs besoins, c'est à dire complets et à jour.

L'approche que nous proposons ici vise d'une part à collecter automatiquement le lexique issu d'un domaine particulier à partir d'un corpus représentatif de ce domaine, et d'autre part à représenter objectivement les acceptions de ce vocabulaire grâce à une méthode mathématique de représentation du sens des mots. La collecte des mots eux-mêmes ne présente pratiquement pas de difficulté, car les systèmes d'analyse morphologique et de lemmatisation sont maintenant très performants, et un traitement semi-automatique des unités lexicales restantes est facilement envisageable. L'établissement du sens au contraire continue de susciter l'intérêt de la recherche actuelle. La méthode que nous suivons consiste à étudier les liens entre unités lexicales pour les projeter grâce à une analyse factorielle des correspondances dans un espace géométrique multidimensionnel pour en visualiser l'organisation sémantique. Cette méthode est une adaptation aux relations syntaxiques du modèles des Atlas sémantiques \citep{Ploux97,PlouxVictorri98}, qui utilise un lien de synonymie entre les mots.

Dans cet article, nous présentons d'abord le modèle mathématique de représentation objective du sens dans son design initial, puis l'adaptation qu'il a subi pour prendre en compte la sémantique lexicale à partir d'un corpus. Ensuite, nous exposons les particularités de notre méthode pour son utilisation dans un domaine spécialisé, ses avantages et contraintes particulières. Enfin, nous replacerons notre approche dans la discipline dont elle est issue, c'est à dire le traitement des connaissances textuelles, et nous présenterons quelques perspectives à cette approche.

\section{Le modèle des Atlas sémantiques}

L'équipe de Sabine Ploux «~Modèles mathématiques et informatiques pour le langage et la perception~» à conçu à l'Institut des Sciences Cognitives du CNRS un modèle tout à fait original de représentation du sens très éloigné du découpage en acceptions propre aux dictionnaires. Ce modèle a été validé par des évaluations lexicologiques \citep{Ploux97,PlouxVictorri98} et psycholinguistiques \citep{RouibahAl01,Maslov04}. Il présente entre autres la particularité d'analyser statistiquement des liens établis par des humains entre unités lexicales afin d'établir le plus objectivement possible une carte représentant les différentes tendances de sens pour chacun des mots traités. Les différentes tendances sont exprimées relativement, au travers des mots auquel le mot considéré est relié. Chaque carte permet donc d'avoir une vision intuitive de la richesse sémantique d'un mot considéré.

Le modèle a été originellement conçu pour résoudre les problèmes relatifs à l'utilisation simultanée de l'information synonymique fournie par sept dictionnaires de synonymes (\textit{Guizot, Lafaye, Bailly, Benac, Du Chazeaud, Larousse, Robert}). En effet, la gestion disparate de l'information de sens effectuée par ces dictionnaires les rendait impropres à une mise en commun simple des correspondances synonymiques. De ce fait, \citep{Ploux97} a proposé d'exploiter la mathématique et la statistique à travers la théorie des graphes et l'analyse factorielle des correspondances pour concevoir une méthode qui articule les mots les uns par rapport aux autres sous l'angle du sens. Ce modèle se veut donc objectif dans la mesure où il est fondé sur une analyse statistique de liens entre les mots étudiés pour aboutir à une représentation du sens. Il se veut également intuitif puisque la distribution des sens est présentée dans un espace géométrique multidimensionnel, où chaque mot se voit représenté dans une carte qui lui est propre. Enfin, il manifeste sa différence en présentant les sens d'un même mot non plus selon un découpage strict des différentes acceptions, mais dans un continuum sémantique où la distance entre deux acceptions est fonction de la différence entre les sens qui leur sont associés.

Le modèle se base sur une seule sorte de relation entre les mots, qui permet de constituer un type de graphe particulier dans lequel toutes les unités lexicales sont interconnectées les unes avec les autres. Ce type de graphe extrêmement dense est appelé «~clique~». Comme le lien qui unit les unités lexicales est un lien de synonymie, les unités qui constituent chaque clique sont très étroitement liées d'un point de vue sémantique. L'interconnexion de chaque unité lexicale composant une clique avec plusieurs autres unités permet de déterminer le sens particulier de chaque unité dans la clique considérée. On pourra par exemple retrouver le mot «~type~» dans deux cliques très différentes, au voisinage de «~amant~» ou «~bonhomme~» pour l'une, dans un sens lié au couple, ou avec «~exemple~» ou «~étalon~» pour l'autre, dans une acception de catégorie. La clique constitue dès lors un niveau de granularité de sens plus fin que le mot lui-même, plus fin que l'acception du dictionnaire également, car les cliques sont généralement plus nombreuses que les acceptions, et deux cliques très voisines peuvent ne varier que par une ou deux unités lexicales, et recouvrir des significations qui se confondent.

\begin{figure}[!ht]
\begin{center}
\includegraphics[width=10cm]{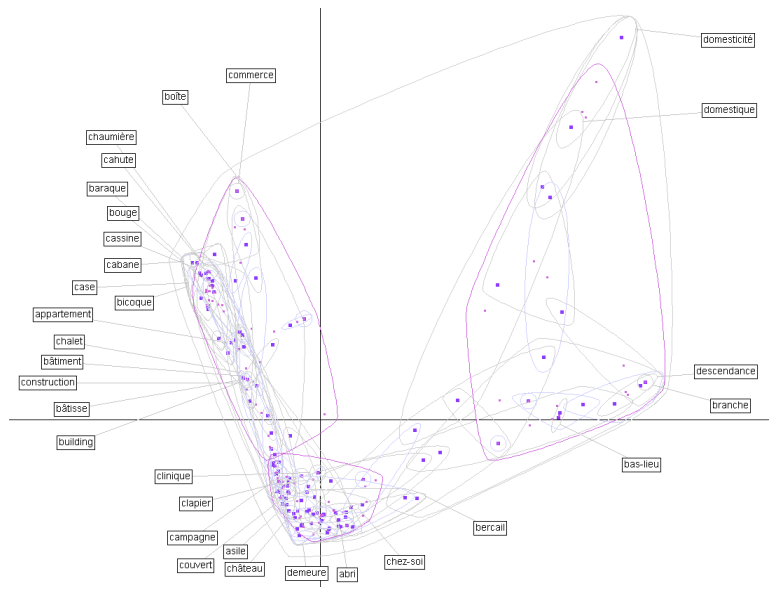}
\caption{Carte sémantique de «~maison~».}\label{maison}
\end{center}
\end{figure}

Une fois les cliques constituées, un traitement statistique appelé analyse factorielle des correspondances est appliqué à chacune des cliques constituées pour une unité donnée. Ce traitement statistique permet de disposer dans un espace géométrique multidimensionnel chacune des cliques dont les coordonnées varieront en fonction de son contenu et de la densité des liens que les différents dictionnaires établissent entre ces unités lexicales. Une projection de cet espace multidimensionnel sur un plan en deux dimensions permet de visualiser les tendances sémantiques du mot considéré, relativement aux synonymes contenus dans les cliques ainsi visualisées. La figure \ref{maison} montre la carte sémantique du mot «~maison~», dont les tendances de sens sont manifestées par des synonymes.

Ce modèle fournit dès lors une ressource sémantique qui représente objectivement des relations d'ordre sémantique -- dont l'établissement n'est pas forcément aussi objectif, puisqu'il est réalisé par les auteurs des différents dictionnaires utilisés. Cette ressource, qui comporte l'information extraite de sept dictionnaires, présente de ce fait une richesse inusitée tout en manifestant une grande cohérence dans son information. Enfin, le continuum sémantique dans lequel chaque tendance sémantique s'inscrit permet de briser la structure classique des dictionnaires dont le découpage en acceptions est souvent arbitraire et trop exclusif.

\section{Atlas sémantique et corpus}

L'excellente capacité du modèle des Atlas sémantiques à décrire objectivement les différents sens d'une unité lexicale grâce aux cliques établis entre cette unité et d'autres qui lui sont particulièrement reliées a amené l'équipe de Sabine Ploux à explorer une autre perspective d'application. En effet, il a semblé intéressant de confronter les qualités indéniables du modèle à un autre type de ressource. Car l'application originelle souffre de son besoin de disposer de plusieurs dictionnaires préexistants, coûteux en temps et en argent, pour construire les cliques. C'est donc logiquement que \citep{JiAl03,JiPloux03} en sont venus à considérer un corpus comme une ressource pertinente pour la construction d'une cartographie sémantique sans dictionnaire.

De fait, un corpus contient intrinsèquement une structure qu'il est aisé de constituer en graphes, à partir desquels peuvent être extraites les cliques lorsque ces graphes contiennent des sommets qui tous sont interconnectés. Ainsi, si la relation entre unités lexicales est l'appartenance à un même contexte, et que le contexte est défini par une fenêtre prédéterminée, toutes les unités appartenant à cette fenêtre sont interconnectées et appartiennent virtuellement à la même clique.

L'utilisation conjointe du modèle des Atlas sémantiques et d'un corpus impose toutefois deux mises en garde. La première concerne l'étendue du corpus. En effet, la richesse de la ressource obtenue est évidemment fonction de la richesse du dictionnaire, c'est-à-dire que seules les unités lexicales représentées dans le corpus pourront être présentes dans la ressource, puisque ce sont ces unités qui servent à construire la ressource. Pour une raison tout aussi évidente, seuls les sens attestés dans le corpus pourront apparaître dans la ressource qui en est issue, puisque l'unité de sens est la clique et que les cliques sont issues de l'analyse du corpus. Le corpus devra donc être suffisamment conséquent non seulement pour contenir un lexique jugé de taille raisonnable, mais également pour que les différents sens de chacune des unités lexicales de ce vocabulaire y soient attestés. La seconde mise en garde découle logiquement de la première. En effet, la taille du corpus ne permet pas d'envisager l'utilisation simple de toutes les cliques concernées, dont le nombre et la diversité seraient sources de bruit. Plusieurs critères de limitation et de contrainte permettent donc d'augmenter la précision du résultat. Notamment, il est possible, pour chaque clique construite pour une unité donnée, de limiter les unités considérées comme pertinentes aux seuls mots dont la fréquence d'apparition dans le contexte de l'unité considérée dépasse un seuil prédéfini. Les contextes rares sont ainsi éliminés. Comme cette contrainte ne suffit pas à limiter suffisamment le bruit, les contextes de ces unités proposées pour la construction des cliques sont eux-mêmes étudiés, de manière à supprimer également les contextes où un co-occurrent fréquent de l'unité considérée se trouve lui-même dans un contexte qui lui est rare. Dans un même ordre d'idée, les mots les plus fréquents du corpus sont également éliminés de la construction des cliques de manière à éviter la présence systématique d'articles, prépositions, auxiliaires et autres mots-outils qui sont porteur d'une sémantique faible et peuvent rarement amener à discriminer les différents sens d'une même unité lexicale.

\begin{figure}[!ht]
\begin{center}
\includegraphics[width=10cm]{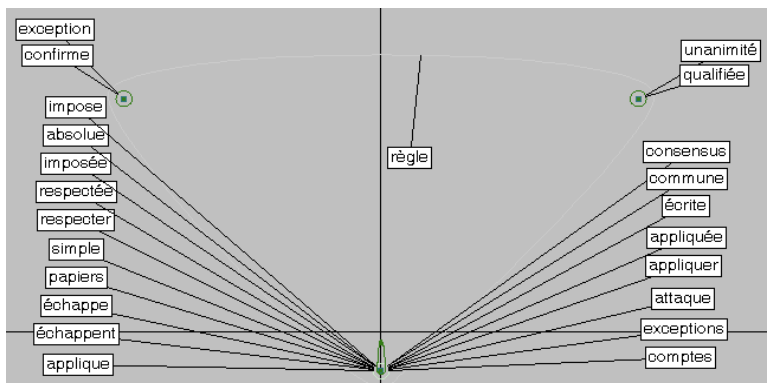}
\caption{Carte sémantique contextuelle de «~règle~».}\label{regle}
\end{center}
\end{figure}

La fenêtre utilisée est soit une fenêtre arbitraire de cinquante mots \citep{JiPloux03}, soit la phrase \citep{JiAl03}. Les cinq cents mots les plus fréquents du corpus sont éliminés du calcul des cliques. Ce sont généralement les unités lexicales qui font partie des cinq pourcents des contextes les plus fréquents qui sont conservées pour construire ces cliques. Une fois ces dernières construites, une analyse factorielle des correspondances semblable à celle utilisée dans le cadre de la synonymie permet de les disposer dans un espace géométrique multidimensionnel, dont la projection dans un plan représente objectivement les diverses tendances d'une unité lexicale donnée. Cependant, à la différence des espaces synonymiques, ce sont les contextes les plus typiques de l'unité considérée dans un sens donné qui permettent d'en distinguer les différents sens\footnote{Un prototype de cette application du modèle des Atlas sémantiques à un corpus est consultable en ligne sur \url{http://dico.isc.cnrs.fr/fr/dico/context/search}. Les corpus utilisés sont le \textit{British National Corpus} pour l'anglais et le corpus du journal \textit{Le Monde} (1997-2002) pour le français.}. La figure \ref{regle} montre la carte sémantique du mot «~règle~» issue l'examen de corpus. Ce sont ici les contextes typiques qui indiquent les différentes tendances de sens.

Dans la perspective d'applications d'ordre psycholinguistique, diverses expériences ont été menées pour tester la validité de cette application du modèle. L'une d'entre elles notamment compare des associations de mots réalisées sur présentation d'un mot-stimulus par des sujets humains et les contextes typiques fournis à ces mêmes mots-stimuli par le modèle \citep{RouibahAl01}. Une autre a montré que la construction de deux ressources distinctes à partir de deux corpus différents (\textit{The MIT Encyclopedia of the Cognitive Sciences} pour l'un et environ 120 \textsl{abstracts} d'articles neuroscientifiques pour l'autre) produit des cartes sémantiques différentes en fonction des habitudes langagières des rédacteurs de l'un ou de l'autre domaine. Si ces expériences tendent à montrer que le modèle est capable de créer une cartographie de la perception du sens chez les rédacteurs d'un corpus ou d'un dictionnaire, elles montrent surtout pour nous, dans le cadre de cet article, une capacité du modèle à représenter automatiquement le sens propre à un corpus. Il est d'ailleurs remarquable que la carte sémantique contextuelle du mot «~règle~» (figure \ref{regle}), obtenue suite à l'examen du corpus \textit{Le Monde 1997-2002}, ne comporte pas toutes les acceptions de ce terme, et omet notamment les sens liés à l'instrument de mesure ou aux menstruations féminines. En effet, ces sens n'apparaissent pas dans le corpus d'origine. Dès lors, il est évident que lorsqu'un corpus spécialisé est utilisé pour établir un dictionnaire sémantique selon la méthode décrite plus haut, la ressource qui en découle est elle-même un dictionnaire de spécialité, dont le lexique et la sémantique sera le fidèle reflet du domaine traité par le corpus, avec les restrictions déjà présentées dans le choix du corpus.

\section{Un corpus spécialisé, une ressource de spécialité}

Mais si le modèle est bien apte à recenser et à représenter les mots et les sens des mots attestés dans un corpus, la méthode mise en {\oe}uvre pour ce faire comporte des lacunes ou des imperfections qu'il est nécessaire de faire disparaître pour augmenter la confiance que l'on peut avoir dans la ressource produite et pour diminuer encore le bruit, qui reste important malgré les techniques de réduction mentionnées plus haut. Les défauts concernent d'abord la faible qualité de la relation entre unités lexicales, ensuite la distinction erronée entre les différentes formes d'une même unité lexicale, enfin la confusion entre des unités distinctes mais homographes. Ces faiblesses de la méthode employée sont sources d'erreurs et nous proposons d'y remédier par une approche mixte qui fait intervenir, en plus des techniques mathématiques et statistiques présentées ci-dessus, quelques traitements d'ordre linguistique.

Notre première proposition concerne l'établissement d'un lien entre les mots plus pertinent que la simple proximité dans une fenêtre arbitraire, et même plus fort que l'appartenance à une même phrase. En effet, l'utilisation statistique d'un seuil de fréquence pour déterminer l'importance d'un contexte dans la méthode d'analyse de corpus présentée ci-dessus, pour efficace qu'elle est, n'en reste pas moins aléatoire et empirique. Nous proposons donc d'exploiter un type de relation qui garantit une interdépendance réelle entre les unités lexicales utilisées comme contextes typiques. Les relations syntaxiques permettent de garantir une appartenance réelle des différentes unités au même contexte. Dès lors, un système d'analyse syntaxique va s'inscrire dans la phase d'examen du corpus, de manière à établir des liens syntaxiques entre unités lexicales et entre têtes de groupes syntaxiques. Toutefois, une application trop stricte de cette forme de contrainte risquerait de faire perdre le bénéfice d'un corpus de taille raisonnable, car certains contextes peuvent apparaître dans ce type de corpus et être considérés comme typiques d'un mot dans un sens donné, sans qu'une relation syntaxique n'unisse au premier chef le terme considéré et son contexte. Une relation syntaxique dite secondaire peut de ce fait être considérée comme pertinente, à condition que le lien primaire ne soit interrompu qu'une seule fois. Ainsi, on pourra considérer que dans les expressions «~décrire un cercle~» et «~décrire un arc de cercle~», le terme «~cercle~» peut également être considéré comme un contexte de «~décrire~». Dans la seconde expression, le lien entre «~décrire~» et «~cercle~» n'est en effet interrompu qu'une seule fois par «~arc~», qui est relié à «~décrire~» et à «~cercle~» par une relation primaire.

Les deuxième et troisième problèmes recouvrent la difficulté que peut éprouver un système automatique pour appréhender la dimension flexionnelle d'un grand nombre d'unités lexicales. En effet, la méthode des contextes proches ne distingue que des séquences de lettres, des graphies, et pas réellement les unités lexicales. De ce fait, les expressions «~il fit des courses~» et «~il fera des courses~» seront traitées séparément, alors que le sens est le même dans les deux cas. De même, chaque forme d'un même nom ou adjectif sera traitée séparément de toutes les autres, ce qui amène à sous-estimer la typicité d'un contexte, ou même à distinguer deux significations différentes pour l'apparition d'un même contexte dans le voisinage d'une même unité lexicale. Il arrive bien sûr que la flexion soit choisie à dessein pour susciter un changement de sens. Ainsi, une expression comme «~faire le trottoir~» n'est pas a priori porteuse de la dimension de travaux publics routiers portée par son pendant «~faire les trottoirs~». Ces changements de sens semblent toutefois plus souvent induits par des variations en morphologie nominale qu'en morphologie verbale. Dès lors, une analyse morphologique complète s'impose comme la solution logique à ces difficultés. Un système d'analyse morphologique automatique permet de distinguer les substantifs des autres catégories grammaticales pour les conserver sous leur forme textuelle, tandis que les autres catégories sont traitées sous leur forme lemmatisée. De la sorte, les différentes formes d'une même unité lexicale peuvent être confondues, tandis que sont distinguées des unités différentes qui présentent une même graphie. Le choix que nous avons fait de ne pas lemmatiser les substantifs est évidemment paramétrable. Par ailleurs, l'analyse morphologique permet également d'éliminer automatiquement les mots-outils dont la sémantique n'est pas pertinente dans ce type d'application. Ainsi, des unités lexicales particulièrement fréquentes ne seront plus supprimées par erreur, tandis que les interjections, prépositions et autres déterminants peu fréquents ne pollueront plus les cartes sémantiques.

Ces différentes propositions nous ont amenés à revoir entièrement les éléments d'information manipulés par le modèle de représentation du sens. En effet, ce sont non seulement les relations entre sommets des graphes (cliques) qui sont modifiées, puisque de relations de proximité, elles sont devenues relations syntaxiques, mais aussi les sommets eux-mêmes, qui ne sont plus des séquences de caractères, mais des lemmes ou des formes de mots. Dès lors, les éléments sur lesquels le modèle peut s'appuyer pour construire les cliques correspondent mieux par leur qualité aux caractéristiques des ressources employées à l'origine pour le valider. La construction des cliques peut de ce fait être calquée sur la méthode utilisée dans le cadre de la construction des Atlas sémantiques synonymiques. En effet, la transformation des données manipulées par le modèle par rapport à la méthode des contextes proches a rendu caduques les contraintes statistiques de limitation du bruit. Ce sont donc des relations syntaxiques primaires ou secondaires entre des lemmes ou formes de mots identifiées qui seront réunies dans les cliques. Comme pour les autres applications, elles sont ensuite disposées dans un espace géométrique multidimensionnel, dont la projection sur un plan permet de visualiser les tendances de sens pour chaque unité lexicale attestée dans le corpus exploité.

Comme dans la méthode des contextes proches, la ressource ainsi produite sera représentative du corpus utilisé pour sa construction. L'ensemble du lexique présent dans le corpus sera représenté dans la ressource sémantique, et tous les sens attestés dans ce corpus trouveront également leur reflet dans les cartes sémantiques. Il nous faut donc à nouveau insister sur la qualité et sur la richesse du corpus utilisé, car la qualité du dictionnaire sémantique sera fonction des facteurs de richesse et d'étendue du corpus exploité. Il faut également noter que plus le corpus utilisé sera représentatif du domaine visé, plus la ressource produite lui sera également adaptée. Le dictionnaire aura des caractéristiques plus générales si des textes plus généraux ou moins ciblés sur le domaine apparaissent dans le corpus.

\section{Ressource spécialisée et traitement de l'information}

Dans une perspective de traitement de l'information textuelle, une ressource spécialisée telle que celle que nous venons de présenter est évidemment d'un apport incalculable. En effet, la représentation des connaissances du domaine abordé représente un atout absolument décisif pour l'identification rapide d'une information recherchée. Pourtant, cet aspect est secondaire par rapport à l'intérêt que ce modèle et l'application que nous en proposons suscite dans les tâches d'extraction de l'information, de recherche d'information, de question-réponse ou de traitement des connaissances. 

En effet, une telle ressource peut apporter la représentation exacte des connaissances contenues dans la base textuelle qu'elle doit aider à interroger si cette base textuelle est utilisée comme corpus spécialisé pour construire la ressource elle-même. La ressource de spécialité est donc spécialisée non seulement dans le domaine du corpus utilisé, mais également ciblé sur le corpus lui-même. De plus, dans le cadre d'une application du modèle au traitement des connaissances d'une base textuelle, nous effectuons une indexation des contenus. Dès lors, une correspondance directe existe entre une tendance de sens représentée sur une carte sémantique pour une unité lexicale et les contextes qui sont à l'origine de cette tendance. De ce fait, il est particulièrement aisé de retrouver toutes les occurrences d'une information dans la base textuelle, par opposition aux occurrences d'une unité lexicale. De la sorte, la précision et la pertinence de l'information obtenue sont pratiquement assurées.

\section{Conclusion}

À travers son application à la relation de proximité textuelle, le modèle des Atlas sémantiques a montré son intérêt et ses grandes qualités dans la description du sens lexical par l'analyse d'une ressource différente des dictionnaires originels. Son aptitude à refléter l'information sémantique contenue dans un corpus textuel transparaît clairement, et de ce fait, il est naturel de s'y intéresser dans le cadre du traitement de l'information textuelle en domaine restreint. Toutefois, si le modèle est bien validé par différentes expériences et évaluations, sa mise en {\oe}uvre souffre de plusieurs imperfections, essentiellement liées à une méthode qui ne prend pas en compte la caractéristique langagière des textes.

Nous proposons donc une nouvelle méthode de mise en {\oe}uvre du modèle qui, utilisant des analyseurs morphologique et syntaxique, amène l'examen du corpus utilisé à un statut de véritable analyse. Le modèle exploite dès lors un corpus composé de termes et plus de séquences de lettres ; les liens utilisés pour composer l'unité de sens, c'est-à-dire la clique, ne sont plus un simple lien de proximité mais cette approche tient compte de la structure du texte à travers les relations syntaxiques. De ce fait, si le modèle reste strictement identique et conserve de ce fait ses qualités -- en particulier son objectivité mathématique --,  les données qu'il manipule sont résolument d'ordre linguistique, et de ce fait prennent en compte la nature même du texte.

La ressource obtenue est un dictionnaire sémantique qui est l'exacte description du vocabulaire contenu dans le corpus utilisé, tant du point de vue lexical que sémantique. Outre cela, ce dictionnaire sémantique est un excellent outil pour naviguer à travers l'information du corpus à partir duquel il a été créé, car une simple indexation des contextes permet d'avoir accès directement aux textes analysés pour le concevoir. En effet, si ce dictionnaire est bien spécialisé dans le domaine dont le corpus utilisé est représentatif, il l'est également dans le corpus lui-même, dont il est le reflet. 

Par ailleurs, d'autres utilisations de cette méthode de mise en {\oe}uvre du modèle peuvent être envisagées. Par exemple, la conception de différentes ressources à partir de corpus similaires, mais propres à des périodes différentes, devrait permettre une vision de la langue en diachronie, en comparant entre elles les différentes cartes d'une même unité lexicale. La construction d'un dictionnaire plus classique pourrait également en être facilitée. En effet, on dispose, en fonction du corpus utilisé, de l'ensemble du vocabulaire pour une langue, ainsi que  des sens existants : tous les sens attestés, mais seulement les sens attestés. De plus, pour chacune des vedettes, on dispose directement d'exemples pertinents et réels, propres à chacun des sens décrits. Enfin, grâce aux modifications que nous avons proposées, diverses informations qui ne sont pas sémantiques sont également disponibles, comme la catégorie grammaticale de la vedette ou divers schémas syntaxiques possibles et bien réels. Comme on peut le voir, la richesse et les qualités du modèle permettent d'envisager bien des perspectives à ses applications.

\section*{Remerciements}

La recherche présentée dans cet article a été financée par le Programme TCAN (Traitement des Connaissances, Apprentissage et NTIC) du CNRS. Nous tenons à adresser les plus vifs remerciements à Hyungsuk Ji, concepteur de l'application des Atlas sémantiques en contexte, sans qui notre recherche n'aurait pu aboutir.

\bibliographystyle{apalike-fr}
% %\bibliographystyle{biblio-fr.bst}
% %\bibliographystyle{abbrv-fr.bst}
% %\bibliographystyle{abbrvnat-fr.bst}
% %\bibliographystyle{alpha-fr.bst}
% %\bibliographystyle{frcomplet.bst}
% %\bibliographystyle{ieeetr-fr.bst}
% %\bibliographystyle{plain-fr.bst}
% %\bibliographystyle{plainnat-fr.bst}
% %\bibliographystyle{siam-fr.bst}
% %\bibliographystyle{unsrt-fr.bst}
% %\bibliographystyle{unsrtnat-fr.bst}

% \bibliography{../../Biblio/bj}
\bibliography{bj.bib}

\end{document}